\documentclass[prl,twocolumn,superscriptaddress,showpacs,longbibliography]{revtex4-1}

\usepackage{mathrsfs}
\usepackage{amssymb, amsbsy, amsmath, latexsym, dsfont, array, layout,
graphicx,mathrsfs,color,ulem,bm}

\begin{document}

\title{Parametric resonance of a Bose-Einstein condensate in a ring trap with periodically driven interactions}
\author{Chen-Xi Zhu}
\affiliation{Key Laboratory of Quantum Information, University of Science and Technology of China, Chinese Academy of Sciences, Hefei, Anhui, 230026, China}
\affiliation{Synergetic Innovation Center of Quantum Information and Quantum Physics, University of Science and Technology of China, Hefei, Anhui 230026, China}
\author{Wei Yi}
\email{wyiz@ustc.edu.cn}
\affiliation{Key Laboratory of Quantum Information, University of Science and Technology of China, Chinese Academy of Sciences, Hefei, Anhui, 230026, China}
\affiliation{Synergetic Innovation Center of Quantum Information and Quantum Physics, University of Science and Technology of China, Hefei, Anhui 230026, China}
\author{Guang-Can Guo}
\affiliation{Key Laboratory of Quantum Information, University of Science and Technology of China, Chinese Academy of Sciences, Hefei, Anhui, 230026, China}
\affiliation{Synergetic Innovation Center of Quantum Information and Quantum Physics, University of Science and Technology of China, Hefei, Anhui 230026, China}
\author{Zheng-Wei Zhou}
\email{zwzhou@ustc.edu.cn}
\affiliation{Key Laboratory of Quantum Information, University of Science and Technology of China, Chinese Academy of Sciences, Hefei, Anhui, 230026, China}
\affiliation{Synergetic Innovation Center of Quantum Information and Quantum Physics, University of Science and Technology of China, Hefei, Anhui 230026, China}

\begin{abstract}
We study the instability of a ring Bose-Einstein condensate under a periodic modulation of inter-atomic interactions. The condensate exhibits temporal and spatial patterns induced by the parametric resonance, which can be characterized by Bogoliubov quasi-particle excitations in the Floquet basis. As the ring geometry significantly limits the number of excitable Bogoliubov modes, we are able to capture the non-linear dynamics of the condensate using a three-mode model. We further demonstrate the robustness of the temporal and spatial patterns against disorder, which are attributed to the mode-locking mechanism under the ring geometry. Our results can be observed in cold atomic systems and are also relevant to physical systems described by the non-linear Schr\"odinger's equation.
\end{abstract}


\maketitle

A uniform system under periodic parametric modulation in time can become unstable and develop a spatial pattern~\cite{faraday_peculiar_1831}. During such a process, known as the parametric resonance~\cite{prbook}, the spontaneous breaking of the spatial translational symmetry is typically accompanied by the breaking of discrete temporal translational symmetry, in the sense that the system acquires a periodic time evolution at different frequencies as that of the parametric driving~\cite{faraday_peculiar_1831,prbook}. Whereas the condition for the occurrence of the parametric resonance is determined by intrinsic properties of a system, the phenomena have been experimentally observed in systems ranging from liquids~\cite{faraday_peculiar_1831,pr3} and solid-state configurations~\cite{pr1,pr2,pr6}, to cold atomic gases in oscillating traps~\cite{engels_observation_2007}. Parametric resonance and the ensuing dynamics have also been under extensive theoretical study in a variety of distinct configurations~\cite{pr4,pr5,pr7,dalfovo,prbec1,prbec2,prbec3,prbec4,staliunas_faraday_2004,nicolin_faraday_2007,staliunas_faraday_2002,suppr1,abdullaev_modulational_1997,kumar_dynamics_2008,abdullaev_collective_2004,conforti_heteroclinic_2016}.
Further, it has been shown that parametric resonance can be suppressed by introducing space- and time-modulated potentials~\cite{suppr2,suppr3}.
In most of these previous studies, the geometry of the system and the boundary condition typically play a minor role.
However, since the system geometry as well as the boundary condition have direct impacts on the form and the availability of the elementary quasi-particle excitations, engineering the system geometry is a promising route toward a better understanding and control of the dynamic instability following a parametric resonance.

In this work, we study the dynamic instability of a quasi-one-dimensional Bose-Einstein condensate in a ring trap. Under a small temporal periodic modulation of the inter-atomic $s$-wave interactions, the uniform condensate can become unstable against temporal and spatial patterns due to the parametric resonance. We clarify the onset of the resonance and the ensuing dynamics by examining the Bogoliubov quasi-particle excitations in the Floquet basis, and find that the dynamics is dominated by a very limited number of Bogoliubov modes under the ring geometry. The system geometry thus not only significantly impacts the resonance condition, but also suppresses higher-order excitations across multiple Floquet bands. This enables us to reproduce the non-linear dynamics using a simplified three-mode model~\cite{conforti_heteroclinic_2016}.

Based on the three-mode model, we find that the overall time evolution is quasi-periodic with rich phase-space contours. At short times, the condensate undergoes the so-called $1/2$ subharmonic resonance, where the condensate oscillates with twice the period of the driving frequency.
While similar subharmonic resonances exist in optical fibres with modulations in the mass term~\cite{conforti_heteroclinic_2016}, in our case, such a behavior can be explained by considering the scattering of condensate atoms into Bogoliubov modes of two adjacent Floquet bands, where the wave vectors of these Bogoliubov modes further determine the spatial pattern. The amplitude of the fast oscillatory dynamics is subject to a slow periodic modulation, due to the non-linearity of the system. Most interestingly, as a result of the mode-locking mechanism enforced by the ring geometry, both the short-time and the spatial oscillatory patterns are robust against disorder. We also discuss the gradual breakdown of the mode-locking mechanism as the temporal modulation increases.

\begin{figure}[tbp]
\includegraphics[width=6cm]{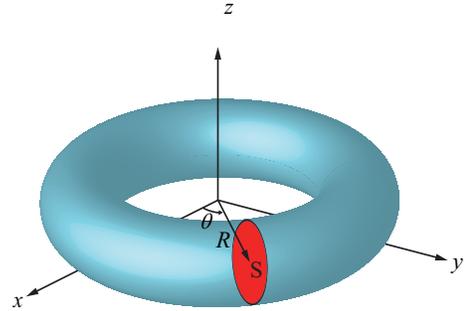}
\caption{Illustration of the ring-trap geometry. See text for detailed discussions.}
\label{fig:fig1}
\end{figure}

{\it Model and Gross-Pitaevskii equation:---}
As illustrated in Fig.~1, we consider a Bose-Einstein condensate in a ring-shaped toroidal trap of radius $R$ and cross section $S$. The condensate is tightly confined in the poloidal direction, and is uniform in the toroidal direction. The parametric resonance is introduced by periodically modulating the $s$-wave interaction strength of the atoms, which can be achieved, for example, by implementing an oscillating magnetic field near a magnetic Feshbach resonance~\cite{chin_feshbach_2010}. The dimensionless Hamiltonian of the system can be written as ($\hbar=1$)
\begin{equation}
\hat{H}=\int_{0}^{2\pi}d\theta\left[-\hat{\psi}^{\dagger}(\theta)\frac{\partial^{2}}{\partial\theta^{2}} \hat{\psi}(\theta)+\frac{U(t)}{2}\hat{\psi}^{\dagger}(\theta)\hat{\psi}^{\dagger} (\theta)\hat{\psi}(\theta)\hat{\psi}(\theta)\right], \label{eq:hamiltonian1}
\end{equation}
where $\hat{\psi}$ and $\hat{\psi}^{\dag}$ are the field operators for the condensate, and $\theta$ is the azimuthal angle. The time-modulated interaction rate $U(t)=U_0[1+\alpha \cos(\omega t)]$, where $U_0$ is the dimensionless interaction rate in the absence of time modulation, and we assume $\alpha\ll 1$. Note that $U_0$ can be related to the $s$-wave scattering length $a_s$ through $U_0=8\pi a_s R/S$. In Eq.~(\ref{eq:hamiltonian1}) and throughout the work, we choose the unit of length and energy to be $R$ and $(2mR^{2})^{-1}$, respectively. We also define $t_0=2mR^{2}$, which serves as the unit of time.

The dynamics of the condensate is determined by the Gross-Pitaevskii (GP) equation
\begin{equation}
i\frac{\partial}{\partial t}\psi=\left[-\frac{\partial^{2}}{\partial\theta^{2}}+U(t)N|\psi|^{2}\right]\psi,
\label{eqn:qpt}
\end{equation}
where $N$ is the total number of atoms, and the mean-field wave function $\psi=\langle \hat{\psi}\rangle$, with the normalization condition $\int_{0}^{2\pi}|\psi|^{2}d\theta=1$. In the absence of interaction modulation, the ground state of the condensate is uniform and stable, so long as $\gamma=U_0 N/2\pi>-1/2$. When $\gamma<-1/2$, the ground state supports a soliton solution~\cite{kanamoto_quantum_2003}. Throughout this work, we will focus on the simpler case of $\gamma>-1/2$.

\begin{figure}[tbh]
\includegraphics[width=8cm]{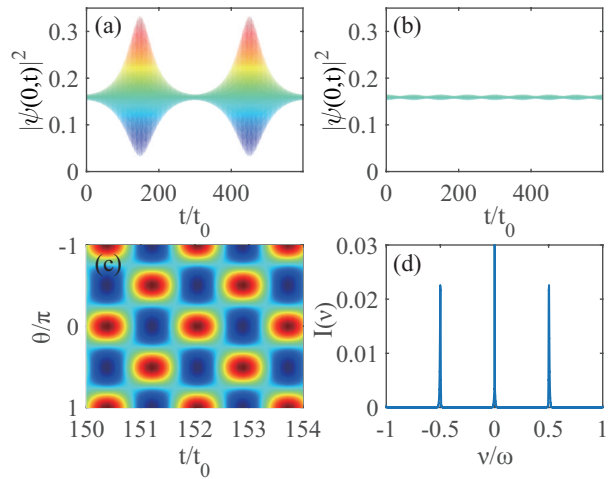}
\caption{(a) Response of the condensate density $|\psi(0,t)|^2$ in the resonance regime under a driving frequency $\omega=7.48$. (b) Response of the condensate density in the non-resonant regime under a driving frequency $\omega=7.38$. (c) The spatial-temporal patterns in the resonance regime with $\omega=7.48$. The spatial pattern here has a period of $\pi$, which corresponds to a wave vector $k=2$. (d) Fourier spectrum of the condensate density $I(\nu)=\int{|\psi(0,t)|^2e^{-i\nu t}dt}$ under a driving frequency $\omega=2\epsilon_{k}$.
In all panels,  $\alpha=0.2$, $U_{0}=-\pi/2$.}
\label{fig:fig2}
\end{figure}

We numerically study the dynamic response of the condensate density to the external modulation by evolving the GP equation, starting from a uniform condensate. When the frequency of the time modulation is within a narrow resonance region, the originally uniform condensate becomes unstable toward the formation of spatial and temporal patterns (see Fig.~\ref{fig:fig2}(a)(c)). This is in sharp contrast to the system response at most modulation frequencies, where condensate would be only slightly perturbed (see Fig.~\ref{fig:fig2}(b)). We note that due to the ring geometry, the dimensionless wave vector associated with the spatial pattern is an integer.
In the resonance region, the dynamic response of the condensate is quasi-periodic, which features fast oscillations enveloped by a periodic amplitude modulation at a much longer time scale. The Fourier spectrum of the condensate density features sharp peaks at $\pm \omega/2$ (see Fig.~\ref{fig:fig2}(d)), which corresponds to a $1/2$ subharmonic oscillation.

{\it Floquet analysis of the quasi-particle spectrum:---}
To understand the dynamic response of the system following the parametric resonance, we study the quasi-particle excitation of the condensate. Following the Bogoliubov theory, we expand the condensate wave function as $\psi=\psi_0+\delta\psi$, where ${\psi_{0}=\frac{1}{\sqrt{2\pi}}\exp{[-i\mu t-i\frac{\alpha U_0N}{2\pi\omega}\sin{(\omega t)}]}}$ is the solution of the GP equation Eq.~(\ref{eqn:qpt}) in the homogeneous case. As we focus on the case of small perturbation with $\alpha\ll 1$, we only keep terms up to linear order in $\delta\psi$, which satisfies
\begin{align}
i\frac{\partial}{\partial t}\delta\psi =&\left[-\frac{\partial^{2}}{\partial\theta^{2}}+2U(t)N|\psi_{0}|^{2}\right]\delta\psi+U(t)N\psi_{0}^{2}\delta\psi^{*}.
\label{eqn:lgp}
\end{align}
We write $\delta\psi$ in the Floquet basis as $\delta\psi(\theta,t)=\psi_0[u(\theta,t)e^{-i\epsilon t}-v^{*}(\theta,t)e^{i\epsilon t}]$, where $u(\theta,t)=\sum_{m}u_{m}(\theta)e^{im\omega t}$ and $v(\theta,t)=\sum_{m}v_{m}(\theta)e^{im\omega t}$. Here $m\in\mathbb{Z}$ is the index for the Floquet bands. The quasi-particle spectrum $\epsilon$ and the Bogoliubov coefficients $\{u_m(\theta),v_m(\theta)\}$ can be solved by diagonalizing the coefficient matrix in the Floquet basis~\cite{tozzo_stability_2005,shirley_solution_1965}.

Instead of numerically diagonalizing the coefficient matrix, we perform Bogoliubov transformation associated with the time-independent case to the coefficient matrix in momentum space. The diagonal elements in the $m$-th Floquet band then become $\pm\epsilon_p+m\omega$, where $\epsilon_p=\sqrt{(p^2+\gamma)^2-\gamma^2}$ and $p$ is the dimensionless momentum. The off-diagonal elements, which couple different Floquet bands, are proportional to $\alpha$ and hence small when the modulation is weak. From the full GP-equation calculations, we find that under the parametric resonance, the dimensionless wave vector $k$ associated with the emergent spatial pattern satisfies $\omega\approx 2\epsilon_k$. This motivates us to focus on the momentum space close to $k$. In this regime, the energy gap within a given Floquet band is much larger than the energy gap between two neighboring bands. It is then sufficient to consider pair-wise the inter-band coupling between the $m$-th and the $(m-1)$-th Floquet bands. Under such an approximation, we derive the dressed quasi-particle spectra from~\cite{supp_material}
\begin{align}
\begin{bmatrix}-\epsilon_{p}+m\omega & \frac{1}{2}(r_{p}+s_{p})^2\alpha\gamma\\
-\frac{1}{2}(r_{p}+s_{p})^2\alpha\gamma & \epsilon_{p}+(m-1)\omega
\end{bmatrix}
\begin{bmatrix}
v_{m}\\u_{m-1}
\end{bmatrix}
=\epsilon\begin{bmatrix}
v_m\\u_{m-1}
\end{bmatrix},\label{eqn:matrix}
\end{align}
where the Bogoliubov modes in the $m$-th and the $(m-1)$-th Floquet bands are coupled. Here, $[r_p,s_p]^{T}$ is the eigenvector in the $\alpha=0$ case with the eigenvalue $\epsilon_p$.

The quasi-particle spectra is then given by
\begin{equation}
\epsilon=(m-\frac{1}{2})\omega\pm \frac{i}{2}\sqrt{(r_p+s_p)^{4}\alpha^{2}\gamma^{2}-(\omega-2\epsilon_{p})^{2}},\label{eq:energy}
\end{equation}
where it is apparent that the quasi-particle energy acquires an imaginary part when $|\omega-2\epsilon_{p}|<(r_p+s_p)^2\alpha|\gamma|$. This corresponds to the dynamic instability of the system. The parametric resonance occurs when a discrete $k$ under the ring geometry happens to be lying in the narrow region $|\omega-2\epsilon_{p}|<(r_p+s_p)^2\alpha|\gamma|$ for any given modulation frequency $\omega$.
This requirement not only significantly limits the available resonance region, but also suppresses higher-order scattering processes with $n\omega\approx 2\epsilon_p$ ($n=2,3,...$). We note that this is quite different from previous studies of the parametric resonance.

\begin{figure}
\includegraphics[width=7cm]{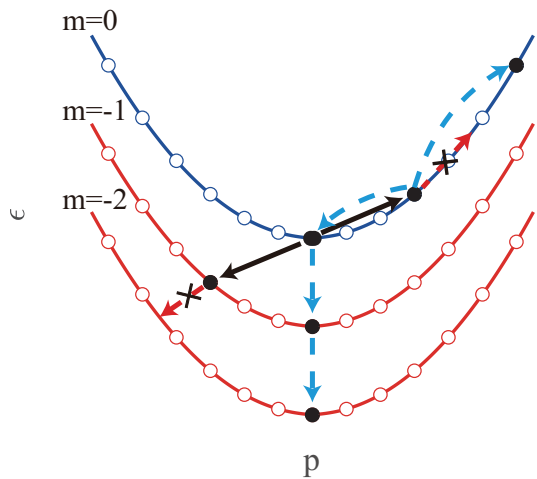}
\caption{Schematics on the scattering processes in a parametric resonance.
Solid curves represent Bologliubov spectra in different Floquet bands with indices $m=0$ (blue) and $m=-1,-2$ (red). While circles on the
spectra represent available states under the ring geometry, the filled and open circles correspond, respectively, to the occupied
and empty states in the scattering process.
The black arrows indicate the dominant scattering process in a parametric resonance. The blue dashed arrows indicate higher-order scattering processes, which only occur when $\alpha$ becomes appreciable. The red arrows indicate higher-order scattering processes which are forbidden by the ring geometry.
}
\label{fig:fig3}
\end{figure}

The condition for parametric resonance can be understood in terms of inter-band phonon scattering between different quasi-particle branches. As illustrated in Fig.~\ref{fig:fig3}, a pair of condensed atoms in the $m=0$ Floquet band are scattered out of the condensate and into quasi-particle states with opposite momenta in Floquet bands $m=0$ and $m=-1$. As a result of the quasi-particle-pair excitation, a standing wave should emerge in the condensate, whose wave vector is determined by the discrete wave vector $k$ satisfying $|\omega-2\epsilon_{k}|<(r_k+s_k)^2\alpha|\gamma|$. Under such a mode-locking mechanism, the system develops temporal oscillations with frequencies sharply centered at $-\omega/2$.
These analysis are in excellent agreement with the numerical results in Fig.~\ref{fig:fig2}.

\begin{figure}
\includegraphics[width=8cm]{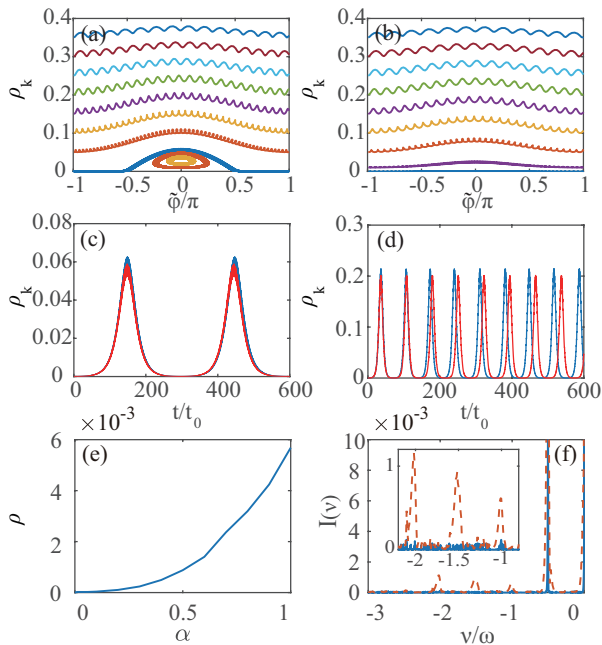}
\caption{(a)(b) Phase-plane contours of the dynamics under the three-mode approximation,
with the parameters $\alpha=0.2$, and driving frequencies (a) $\omega=7.48$ and (b) $\omega=7.38$. In both cases, the resonant wave vector is $k=2$.
(c)(d) Comparison of $\rho_k$ between results from the GP equation Eq.~(\ref{eqn:qpt})(blue) and under the three-mode approximation (red), with the parameters (c) $\alpha=0.2$ and (d) $\alpha=1$. (e) Population of modes other than $p=0,\pm 2$ as a function of $\alpha$, calculated using the GP equation. Here $\rho=1-\rho_0-2\rho_k$.
(f) Fourier spectrum of the condensate density $I(\nu)$ with $\omega=7.48$, $\alpha=0.2$ (blue) and $\alpha=1$ (red). In all panels, $U_0 = -\pi/2$. In (c)(d)(e)(f), $\omega=7.48$.}
\label{fig:fig4}
\end{figure}

{\it Three-mode model and long-time dynamics:---}
The mode-locking mechanism of the ring geometry suggests that we should be able to capture the dominant dynamics of the system using a simplifed three-mode model, which only takes into account the condensate mode $p=0$ and the two quasi-particle modes $p=\pm k$.

We start from the second-quantized Hamiltonian
\begin{equation}
H=\sum_{p}\epsilon_{p}^{0}a_{p}^{\dagger}a_{p}+\frac{U(t)}{2V}\sum_{p,p',q}a_{p+q}^{\dagger}a_{p'-q}^{\dagger}a_{p'}a_{p},
\end{equation}
where $a_p$ ($a^{\dag}_p$) is the annihilation (creation) operator for atoms with momentum $p$. Under the three-mode approximation discussed above, we only retain operators with $p=0,\pm k$. Applying the Heisenberg equations of motion for the operators $\{\hat{a}_0,\hat{a}_k,\hat{a}_{-k}\}$, and taking the mean-field approximation $\hat{a}_{j}=\sqrt{n_{j}}e^{i\theta_{j}(t)}$ ($j=0,\pm k$), we have~\cite{supp_material}
\begin{align}
\dot{\rho}_{k} & =2U'(1-2\rho_{k})\rho_{k}\sin\varphi,\label{eq:rho},\\
\dot{\varphi} & =2k^{2}+2U'(1-3\rho_{k})+2U'(1-4\rho_{k})\cos\varphi.\label{eq:phi}
\end{align}
Here, $U'=UN/2\pi$, $\rho_{j}=n_j/N$, $\varphi=2(\theta_0-\theta_k)$, and $N=n_0+2n_k$ is the total atom number. Note that we have assumed $n_k=n_{-k}$, $\theta_k=\theta_{-k}$, due to the symmetry of the scattering process.

The non-linear equations Eqs.~(\ref{eq:rho}) and (\ref{eq:phi}) resemble those characterizing the non-rigid-pendulum motion of the dynamics of spinor condensates~\cite{spinor,zhou_spin_2010,liu_quantum_2009}. Similar to the practice therein, we characterize the system dynamics by calculating the phase-plane contour. From Eq.~(\ref{eq:phi}), it is apparent that $\varphi$ undergoes rapid oscillatory motion with a frequency $2k^{2}+\frac{U_0N}{\pi}(1-3\langle\rho_{k}\rangle_t)$, which is essentially the fast oscillation we see in Fig.~\ref{fig:fig2}(c). We therefore define $\tilde{\varphi}=\varphi-\omega t$, and characterize the long-term dynamics of the system with contour plots on the $\rho_k$-$\tilde{\varphi}$ plane.
As shown in in Fig.~\ref{fig:fig4}(a), the dynamics can be roughly classified as a self-trapping phase, with closed contours and `localized' ($\tilde\varphi$); and a running phase, with ($\tilde\varphi$) taking values from $-\pi$ to $\pi$. Whereas the long-time dynamics is characterized by the system going around a given contour in Fig.~\ref{fig:fig4}(a), the occurrence of the parametric resonance is indicated by the existence of the self-trapping phase in the contour plot.

Our analysis so far is based on the perturbation theory, assuming small temporal modulations. As the modulation amplitude $\alpha$ increases, the mode-locking mechanism would eventually breakdown. To understand the process, we study the response of the system with larger $\alpha$ by solving the GP equation (\ref{eq:cos(qx)}). As shown in Fig.~\ref{fig:fig4}(c)(d), the three-mode approximation faithfully reproduces results from the GP equation at a small $\alpha$, but significantly deviates from the solution from GP equation for a larger $\alpha$, particularly at long times. This is due to the occupation of modes other than $0,\pm k$, induced by higher-order scattering processes under the large modulation amplitude.
In Fig.~\ref{fig:fig4}(e), we show the population of these other modes as a function of $\alpha$.
As $\alpha$ increases, the occupation of modes other than $0,\pm k$ increases only slightly. Their impact on the overall dynamics as illustrated in Fig.~~\ref{fig:fig4}(d) therefore illustrates the highly non-linear nature of the system.

The impact of these other excitation mode is reflected in the Fourier spectrum of the condensate density. In Fig.~\ref{fig:fig4}(f), we show the spectrum at $\alpha=1$. Besides the parametric resonance peak near $\nu/\omega=-0.5$, we identify several smaller peaks at higher frequencies, which correspond to higher-order inter-band scattering processes. As illustrated in Fig.~\ref{fig:fig3}, these higher-order processes give rise to small occupations at various momenta of different Floquet bands. The interference between these modes then contribute to the peaks in the Fourier spectrum. For example, the leading-order contribution to the peak near $\nu/\omega=-2$ comes from the interference between the condensate at $k=0$ and excitations in the $m=-2$ band at $k=0$.

\begin{figure}[tbp]
\includegraphics[width=8cm]{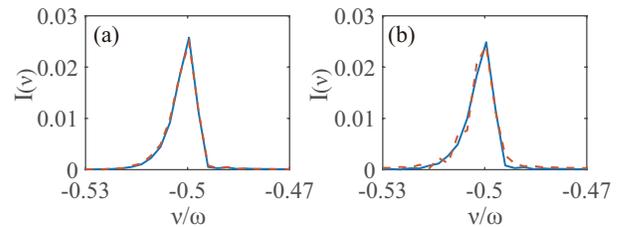}
\caption{Fourier spectra of the condensate density $I(\nu)$ near $\nu=-\omega/2$
for: (a) $V_{d}=0$ and $q=0$ (blue solid), $q=0.2$ (red dashed); (b) $q=0$ and $V_{d}=0$(blue solid), $V_{d}=0.02$ (red dashed). The results are similar for $I(\nu)$ near $\nu =\omega/2$.}
\label{fig:fig5}
\end{figure}

{\it Robustness of the mode-locking mechanism:---}
A unique feature of our system is the robustness of both the temporal and spatial patterns against disorder. This is due to the mode-locking mechanism under the ring geometry. To demonstrate this, we introduce a disordered GP equation
\begin{equation}
i\partial_{t}\psi=[-\partial_{\theta}^{2}+V(\theta)+U(1+\alpha cos(\omega t)cos(q\theta))|\psi|^{2}]\psi, \label{eq:cos(qx)}
\end{equation}
where $V(\theta)=V_d g(\theta)$ is a random potential in space with its strength $V_d$, and $g(\theta)$ is the normal distribution. Another source of disorder comes from the term $\cos(q\theta)$ in the interaction, which gives a finite momentum transfer $q$ to the scattering processes in Fig.~\ref{fig:fig2}. We numerically evolve the full GP equation (\ref{eq:cos(qx)}), and plot the Fourier spectrum of the resulting dynamics. As illustrated in Fig.~\ref{fig:fig5}(a)(b), the dominant peak at $-\omega/2$ is robust against both types of disorder. On the other hand, as the spatial pattern is dictated by wave vectors of the Bogoliubov modes responsible for the fast oscillatory dynamics, it is also robust against both types of disorder. In contrast, the long-time slow dynamics, being a result of non-linearity, is not robust against disorder.

{\it Conclusions:---}
We have studied the parametric resonance of a Bose-Einstein condensate in a ring geometry. As a result of periodic modulation of the inter-atomic interaction strength, the condensate spontaneously breaks spatial and discrete-time translational symmetry and develops spatial and temporal patterns. A unique feature of the ring geometry is the mode-selection mechanism, which significantly suppresses higher-order processes in the dynamics and makes the system response robust to disorders. Such a behavior is similar to the recently proposed Floquet time crystals~\cite{sacha_modeling_2015,wilczek_quantum_2012}, where the spontaneous breaking of discrete time-translational symmetry is also robust against disorder~\cite{yao_discrete_2017,sacha_quantum_2018}. Our results can be observed in cold atomic systems and are relevant to physical systems described by the non-linear Schr\"odinger's equation.

{\it Acknowledgements:---}
We thank Ming Gong for useful discussions. This work is supported by the National Natural Science Foundation of China (Grant Nos. 11522545, 11574294) and the National Key Research and Development Plan (Grant Nos. 2016YFA0301700, 2016YFA0302700, 2017YFA0304100).

\end{document}